# Infrared spectrum and intermolecular potential energy surface of the CO-$O_2$ dimer

A.J. Barclay,[a] A.R.W. McKellar,[b] N. Moazzen-Ahmadi,*[a]
Richard Dawes,[c] Xiao-Gang Wang,[d] and Tucker Carrington Jr.[d]

**Abstract**

Only a few weakly-bound complexes containing the $O_2$ molecule have been characterized by high resolution spectroscopy, no doubt due to the complications added by the oxygen molecule's unpaired electron spin. Here we report an extensive infrared spectrum of CO-$O_2$, observed in the CO fundamental band region using a tunable quantum cascade laser to probe a pulsed supersonic jet expansion. The rotational energy level pattern derived from the spectrum consists of stacks of levels characterized by the total angular momentum, $J$, and its projection on the intermolecular axis, $K$. Five such stacks are observed in the ground vibrational state, and ten in the excited state ($v$(CO) = 1). They are divided into two groups, with no observed transitions between groups. The groups correspond to different projections of the $O_2$ electron spin, and correlate with the two lowest rotational states of $O_2$, $(N, J) = (1, 0)$ and $(1, 2)$. The rotational constant of the lowest $K = 0$ stack implies an effective intermolecular separation of 3.82 Å, but this should be interpreted with caution since it ignores possible effects of electron spin. A new high-level 4-dimensional potential energy surface is developed for CO-$O_2$, and rotational energy levels are calculated for this surface, ignoring electron spin. By comparing calculated and observed levels, it is possible to assign detailed quantum labels to the observed level stacks.

---

[a] *Department of Physics and Astronomy, University of Calgary, 2500 University Drive North West, Calgary, Alberta T2N 1N4, Canada*

[b] *National Research Council of Canada, Ottawa, Ontario K1A 0R6, Canada*

[c] *Department of Chemistry, Missouri University of Science and Technology, Rolla, MO 65409-0010, USA*

[d] *Chemistry Department, Queen's University, Kingston, Ontario K7L 3N6, Canada*

# Introduction

Hundreds of weakly-bound van der Waals complexes have now been characterized by high resolution spectroscopy.[1] Relatively few of these involve an open shell constituent (with unpaired electron spin or orbital angular momentum), but there are still many such examples, as described in two review articles.[2,3] A very short but representative list, focusing on microwave and infrared results, includes: Ar-NO,[4,5] Ar-NO$_2$,[6] HF-NO,[7,8] Ar-OH,[9,10] and Ar-HO$_2$.[11] The number of such complexes containing the O$_2$ molecule is however quite limited: (O$_2$)$_2$,[12-14] Ar-O$_2$,[15-17] HF-O$_2$,[18,19] N$_2$O-O$_2$,[20,21] and H$_2$O-O$_2$.[22,23] There is also unpublished work on SO$_2$-O$_2$[24] and OCS-O$_2$.[25]

In the present paper, we study in detail the infrared spectrum of a new radical complex, CO-O$_2$, as observed in the region of the fundamental vibration of carbon monoxide ($\approx$2150 cm$^{-1}$) using a pulsed supersonic slit jet expansion. The dynamics of CO-O$_2$ lie intermediate between the limits of free internal rotation and "normal" semi-rigid molecule behavior. We are exploring new territory for oxygen-containing complexes since the examples given above (HF-O$_2$, N$_2$O-O$_2$) are much closer to the semi-rigid limiting case. Indeed, the published analyses of HF-O$_2$ and N$_2$O-O$_2$ used a Hamiltonian which assumes a fixed angle for O$_2$ relative to the intermolecular axis,[18-20] implying, among other things, that the oxygen O atoms are inequivalent. We believe that this Hamiltonian is not appropriate here because CO-O$_2$ is considerably less rigid structurally than HF- or N$_2$O-O$_2$.

We find that the energy level pattern of CO-O$_2$ consists of various "stacks" which are well characterized by $K$, the projection of the total angular momentum, $J$, on the intermolecular axis. Within each stack, $J = K$, $K + 1$, $K + 2$, etc. Five such stacks are observed in the ground vibrational state, and ten in the excited (v(CO) = 1) state, and they can be divided into two separate groups, with no observed transitions between groups. We believe that these distinct groups correspond to different projections of the O$_2$ electron spin, $S = 1$. Apart from spin effects, the spectrum and energy levels of CO-O$_2$ might be similar to those of CO-N$_2$, which has been studied in detail.[26-34] The approach used to analyze CO-N$_2$ spectra has been simply to fit different stack origins, rotational and centrifugal distortion constants ($\sigma_0$, $B$, $D$, etc.) for each $K$-stack, and we adopt the same approach here for CO-O$_2$. Since there is no satisfactory effective Hamiltonian for CO-N$_2$, we do not expect one for CO-O$_2$ for which spin adds an extra complication.

In order to better understand these results, we also report here a new high-level *ab initio* potential energy surface for CO-O$_2$, together with rovibrational energy levels calculated for this surface. These calculated levels, organized into $K$-stacks, are very helpful in assigning detailed quantum labels for the observed $K$-stacks, even though they do not include the electron spin. The meaningful stack labels turn out to be the projections on the intermolecular axis of the O$_2$ rotation, the CO rotation, and the spin. The sum of these projections is, of course, the $K$-value that characterizes the stack.

The presentation below begins with the observed spectrum and its interpretation in terms of CO-O$_2$ energy levels, without much reference to electron spin. Then we discuss the interpretation of the various observed $K$-stacks in terms of free O$_2$ and CO rotation. Moving to *ab initio* theory, a 4-dimensional potential energy surface is described, and rovibrational energy levels are calculated on this surface. These results then help us assign detailed quantum labels to the observed $K$-stacks. The last section provides further discussion and conclusions, including predictions for the as yet unobserved microwave spectrum of CO-$O_2$.

## The observed spectrum

Spectra were recorded at the University of Calgary as described previously,[34-37] using a pulsed supersonic slit jet apparatus and a Daylight Solutions quantum cascade laser. The expansion mixture contained about 0.1 to 0.3% carbon monoxide plus 0.3 to 0.9% oxygen in helium carrier gas, and the jet backing pressure was 9 atmospheres. Under these conditions, the CO dimer spectrum[37] was observed along with that of CO-$O_2$. Wavenumber calibration was carried out by simultaneously recording signals from a fixed etalon and a reference gas cell containing $N_2O$. Spectral simulation was aided using the PGOPHER software.[38]

### Group 1, levels correlating with ($n(O_2)$, $j(O_2)$) = (1, 0)

The top trace of Fig. 1 shows part of the observed spectrum, with a He + $O_2$ + CO expansion mixture plotted in red and a He + CO mixture plotted in black in front. This helps to distinguish the CO-$O_2$ lines as those which "stick out" in red behind the black CO dimer lines (though the cancellation is not perfect since the effective rotational temperatures are slightly different in the two spectra). A prominent CO-$O_2$ $Q$-branch feature around 2145.5 cm$^{-1}$ is similar to features observed for CO-$N_2$ ($\approx$2146.2 cm$^{-1}$),[26] CO-Ar ($\approx$2145.2 cm$^{-1}$),[39] and CO-Ne ($\approx$2146.4 cm$^{-1}$).[40] By analogy, it was thus easy to assign $P$-, $Q$-, and $R$-branch transitions of a $K = 1 \leftarrow 0$ band of CO-$O_2$, as illustrated by a simulated spectrum in Fig. 1. A mirror-image $K = 0 \leftarrow 1$ band, with its $Q$-branch around 2139.9 cm$^{-1}$ (not shown here) was also easily assigned. With more difficulty, we located a weaker $K = 0 \leftarrow 0$ band which unambiguously involves the same $K = 0$ stacks and is centered at 2142.7 cm$^{-1}$. The corresponding $K = 1 \leftarrow 1$ band could not be clearly detected. All assigned transitions are listed in Tables A-1 and A-2 of the Electronic Supplementary Information.

Guided by the analogy with CO-$N_2$ and CO-Ar, and by the ground state combination differences (energy level separations) already determined above, we located a $K = 2 \leftarrow 1$ band centered around 2149.5 cm$^{-1}$ (an analogous band of CO-$N_2$ lies around 2150.1 cm$^{-1}$).[26] In addition, there were two prominent $K = 0 \leftarrow 0$ bands centered at 2151.8 and 2152.8 cm$^{-1}$ (see Fig. 2). These involved transitions from the already known ground state (v(CO) = 0) $K = 0$ stack to two new excited state (v(CO) = 1) $K = 0$ stacks, which we label 0' and 0". These new $K = 0$ upper states involve large changes in $B$-values (especially the latter) so the $P$- and $R$-branch structures of the bands themselves were not so obvious at first. But the assignments are completely confirmed by ground state combination differences which match those already known for the ground $K = 0$ stack.

All the transitions discussed so far can be explained in terms of about 70 rotational energy levels belonging to two ground state (v(CO) = 0) and five excited state (v(CO) = 1) $K$-stacks, as illustrated in Fig. 3 and listed in Table 1. Note that the two lowest stacks, $K = 0$ and 1, are almost identical for v(CO) = 0 and 1, apart from the difference of 2142.694 cm$^{-1}$ (this represents a vibrational shift of -0.577 cm$^{-1}$, relative to the free CO monomer). Thus we expect the energy level scheme to remain similar in the upper and lower states, and there is no reason to doubt that the higher stacks, $K = 2$, 0', and 0", are also present for v(CO) = 0. They remain unobserved simply because they are almost unpopulated at our experimental temperature of around 2.2 K.

Interestingly, levels of the excited state (v(CO) = 1) $K = 2e$ stack cross those of the $K = 0"$ stack between $J = 5$ and 6. This crossing involves some mixing of states, as shown by the fact that we observed some satellite transitions with $K = 0" \leftarrow 1$ in the region of the "allowed" $K = 2 \leftarrow 1$ band, and with $K = 2 \leftarrow 0$ in the region of the "allowed" $K = 0" \leftarrow 0$ band. Unfortunately, the crossing region around $J = 6$ is where these transitions become too weak to assign reliably, so we only have a partial picture of this level crossing.

### Group 2, levels correlating with ($n(O_2)$, $j(O_2)$) = (1, 2)

The transitions discussed so far explain much, but not all, of the observed CO-$O_2$ spectrum. For example, in the region of Fig. 1 we were able to assign a $K = 3 \leftarrow 2$ band with a Q-branch at about 2145.3 cm$^{-1}$ and an R-branch starting with a strong line at 2145.74 cm$^{-1}$ (see the simulation in Fig. 1). Taking this band together with its mirror-image $K = 2 \leftarrow 3$ band and a weak central $K = 2 \leftarrow 2$ band enabled us to characterize these new $K = 2$ and 3 stacks in both the ground and excited vibrational states (v(CO) = 0 and 1). Further investigation revealed a $K = 1 \leftarrow 2$ band (see Fig. 1), a $K = 4 \leftarrow 3$ band (close to the $K = 2 \leftarrow 1$ band of the previous section), and another $K = 2 \leftarrow 2$ band involving a second excited state stack which we label $K = 2'$ (some transitions of this band are marked with pound signs in Fig. 2). This may seem confusing, but the energy level scheme shown in Fig. 4 and listed in Table 2 should help to clarify the situation. As in the preceding section, the lower stacks, $K = 2$, 3, and 1, are very similar for v(CO) = 0 and 1, and this undoubtedly continues for the higher stacks, $K = 4$ and 2', even though they are not observed for v(CO) = 0.

No transitions were observed which connected the present "group 2" $K$-stacks (Fig. 4, Table 2) with the "group 1" stacks of the previous section (Fig. 3, Table 1). So we can only estimate, based on observed intensities (and assuming similar transition strengths), that group 2 lies roughly 2 cm$^{-1}$ above group 1. As discussed below, we believe these two non-interacting groups of levels correlate with the two lowest rotational levels of $O_2$, namely $(n, j) = (1, 0)$ and $(1, 2)$. The vibrational red shift of -0.577 cm$^{-1}$ is the same for the two groups within experimental error.

### Empirical parameters

The 'experimental' energy levels in Tables 1 and 2 were fitted using the following simple empirical expression,

$$E = \sigma + B\,[J(J+1) - K^2] - D\,[J(J+1) - K^2]^2 + H\,[J(J+1) - K^2]^3$$
$$\pm (1/2)\{b[J(J+1)] + d[J(J+1)]^2 + h[J(J+1)]^3\}, \quad (1)$$

where $\sigma$ is the $K$-state origin, $B$ is the rotational constant, and $D$ and $H$ are centrifugal distortion constants. Parameters $b$, $d$, and $h$ express the splitting of $e$ and $f$ components for $K > 0$ (plus sign for $e$ and minus sign for $f$). We expect that $b = 0$ for $K > 1$, $d = 0$ for $K > 2$, etc. This is the same expression as used previously for CO-$N_2$, facilitating comparison of the two species.[26,28-34] Results of the fits are given in Tables 3 and 4; we omit giving any uncertainties here because in many cases the number of levels fitted is not much larger than the number of parameters (and the Hamiltonian may not be fully appropriate). The rotational constant of the lowest $K = 0$ stack, 0.0772 cm$^{-1}$, may be compared to values of 0.0743 and 0.0708 cm$^{-1}$ for CO-*ortho*N$_2$ and CO-*para*N$_2$, respectively. It implies an effective ground state intermolecular separation of 3.82 Å for CO-O$_2$, but this should be interpreted with caution since it ignores possible effects of electron spin. For comparison, in CO-*ortho*N$_2$ the lowest $K$-stack implies a separation of 4.03 Å, but (as in the present case) there is quite a range of $B$-values among different stacks. The CO dimer in effect has two ground states, one C-bonded with $R \approx 4.4$ Å and the other O-bonded with $R \approx 4.0$ Å.[37]

## Free rotor interpretation

As a starting point, it is useful to think in terms of free rotation for the CO and O$_2$ monomers within the CO-O$_2$ dimer. In its $^3\Sigma_g^-$ ground electronic state, molecular oxygen has a net unpaired electron spin angular momentum of $S = 1$ which couples with the rotational angular momentum, $n(O_2)$, to give total angular momentum, $j(O_2)$. Only odd values of $n(O_2)$ are allowed because of the zero nuclear spin of the O atom and the negative electronic state parity. The lowest allowed rotational level, $n(O_2) = 1$, splits into three spin components, $j(O_2) = 0, 2$, and 1, which have energies of about 0.00, 2.10, and 3.97 cm$^{-1}$, respectively. The next rotational level, $n(O_2) = 3$, similarly has components with $j(O_2) = 2, 4$, and 3, at about 16.24, 16.43, and 18.35 cm$^{-1}$. Meanwhile, the CO molecule in its closed-shell $^1\Sigma^+$ ground electronic state has rotational levels $j(CO)$ ($\equiv n(CO)$) = 0, 1, 2, 3, etc., with energies of about 0.0, 3.85, 11.55, 23.07 cm$^{-1}$, respectively. By summing the O$_2$ and CO energies, we obtain free-rotor energy levels for CO + O$_2$ as shown on the left-hand side of Fig. 5. The levels are coded in red for $j(O_2) = n(O_2) - 1$, blue for $j(O_2) = n(O_2) + 1$, and black for $j(O_2) = n(O_2)$. Each free rotor level can then have a stack of dimer rotational levels built on it, adding energies approximately equal to $B(CO-O_2) \times L(L + 1)$, where $B(CO-O_2) \approx 0.077$ cm$^{-1}$, and $L$ is the quantum number for end-over-end rotation.

The right-hand side of Fig. 5 shows the fitted stack origins for CO-O$_2$ from Tables 3 and 4. Here we use the more complete upper state (v(CO) = 1) data, recalling that the ground state is very similar. Note that the exact values of the stack origins depend on how they are defined (e.g. the $-K^2$ terms in Eq. 1). The relative energies of the two groups (coded here in red and blue) are not exactly known, so we use the previously mentioned approximate separation of 2 cm$^{-1}$ based on relative intensities. The dashed lines in Fig. 5 show our proposed correlation of the observed CO-O$_2$ $K$-stacks with the free rotor levels. Note that group 1 (red) correlates to free rotor levels with $(n(O_2), j(O_2)) = (1, 0)$, and group 2 (blue) correlates to $(n(O_2), j(O_2)) = (1, 2)$. Analogous plots for CO-N$_2$ are shown in Figs. 2 and 3 of Ref. 34.

The intensity in our spectrum derives from the CO vibrational transition moment, so in the free-rotor limit the selection rule is $\Delta j(CO) = \pm 1$. The thin vertical lines on the left hand side of Fig. 5 show these allowed free rotor transitions. The corresponding vertical lines on the right side correspond to the sub-bands actually observed in our spectrum (except of course the observed transitions are between different CO vibrational states (v(CO) = 0 and 1), not within one state as shown in Fig. 5). As expected, these observed CO-O$_2$ bands all correlate with allowed ($\Delta j(CO) = \pm 1$) free rotor transitions. In addition to the eight subbands shown in Fig. 5, we also observed bands with $K = 0 \leftarrow 0$ in group 1, and $2 \leftarrow 2$ in group 2. But they are relatively weak, and the weakness can be explained by the fact that they correlate with $\Delta j(CO) = 0$. Figure 5 emphasizes what was evident in the spectrum, namely that the $K = 1 \leftarrow 0$ band in group 1 is analogous to the $K = 3 \leftarrow 2$ and $1 \leftarrow 2$ bands in group 2. Similarly, the $2 \leftarrow 1$ band in group 1 is analogous to the $4 \leftarrow 3$ band in group 2.

Experiment does not distinguish the $e$ and $f$ spectroscopic parity labels used in Figs. 3, 4, and Tables 1 - 4. We chose the $e$ label for the lowest $K = 0$ stack of group 1 based on the free rotor limit, since the lowest $(n(O_2), j(O_2)) = (1, 0)$ rotational level of the $O_2$ monomer has positive parity,[41] and this determines the other group 1 stacks as shown in Fig. 3. However, the lowest $J = 0$, $K = 0$ level of Ar-$O_2$ was labelled as having negative parity in Fig. 1 of Ref. 16, so it is possible that our $e$ and $f$ labels should be reversed. For group 2 (Fig. 4, Tables 2 and 4), the $e/f$ splittings are mostly small and the labelling is more problematical. But the relative labelling is still well established, except for the $K = 4$ stack where the splittings are somewhat erratic.

## Theory

### 4D Potential Energy Surface

To guide interpretation of the experimental results, a 4D potential energy surface (PES) was constructed, describing the interaction between CO($\Sigma_g^+$) and $O_2$($\Sigma_g^-$), held rigid at their ground vibrational state averaged bond distances (1.12821 and 1.20752 Å respectively). The construction of this PES was used as an illustrative example and described in some detail in a recent review of *ab initio* methods and procedures suitable for use in such applications.[42] To summarize, an automated procedure was used to fit the PES using the Interpolating Moving Least Squares (IMLS) method.[43,44] This approach has been applied previously to numerous van der Waals systems composed of linear fragments: (OCS)$_2$,[45] (CO)$_2$, (CO$_2$)$_2$,[46,47] CO$_2$-CS$_2$,[48] CO-N$_2$, (NNO)$_2$,[46,49] CO$_2$-HCCH,[50] and C$_6$H$^-$-H$_2$.[51] Here, a total of 1932 symmetry unique points were required to achieve an estimated root-mean-square (rms) fitting error below 0.1 cm$^{-1}$. Since the method is interpolative (the fit passes through all included data points) an algorithm is used to estimate the overall fidelity to the surface.[43] Less complete *ab initio* studies of the CO-O$_2$ system have been reported by Grein[52] and by Tashakor et al.[53]

The Molpro electronic structure code was used for all of the calculations reported here.[54] In order to determine an appropriate level of *ab initio* theory suitable for the global PES, a series of benchmarks were performed using the structures of two planar isomers. Shown in Fig. 6, the structures of two planar isomers, denoted *o*-in (global minimum) and *c*-in, were initially located and optimized at the UCCSD(T*)-F12a/VDZ-F12 level (where (T*) indicates scaling of the triples contribution by the ratio of the MP2-F12/MP2 correlation energy, see Molpro manual). The relative energies of the two isomers are sensitive to basis set completeness and core-correlation. Table 1 of Ref. 42 lists the interaction energies for the two isomers as a function of basis set completeness (up to the CBS limit) for the unrestricted explicitly-correlated coupled-cluster method, comparing valence-only and all-electron correlation [UCCSD(T)-F12b/VnZ-F12 and (AE)UCCSD(T)-F12b/CVnZ-F12]. As discussed in Ref. 42, although the effect of correlating the core-electrons could be viewed as significant at particular intermediate basis sizes, at the CBS limits, the valence-only and all-electron correlation calculations both converge to essentially the same relative energies for the two isomers. Thus to make the global PES, CBS limit energies were obtained by extrapolation of valence-only calculations at the UCCSD(T)-F12/VTZ-F12 and UCCSD(T)-F12/VQZ-F12 levels. The extrapolation employed a scheme suggested by Schwenke[55] with a parameter value of $F = 2.06$.

In Fig. 7, the PES is plotted for planar geometries as function of two extended angles which describe the complete 360° rotation of each fragment. For each pair of angles, the center-of-mass distance between fragments is varied to minimize the energy and thus the plot represents the fully relaxed structures of any planar isomers. As seen in the plot, the global minimum $o$-in structure has a very low-energy disrotatory path or channel, connecting to an equivalent structure. The saddle-point of that path is a T-shaped structure with the O-atom of CO pointing to the side of $O_2$. This saddle was reported as a stable minimum by Grein.[52] A slightly higher energy conrotatory path connects the $o$-in isomer to the less stable $c$-in isomer (a structure not reported by Grein). The energies of the fully relaxed $o$-in and $c$-in isomers on the CBS PES are $E = -119.3$ and $E = -112.8$ cm$^{-1}$ respectively. Grein also reports a cross-shaped non-planar local minimum, corresponding to which a similar structure is found as a stable minimum on our PES ($E = -116.8$ cm$^{-1}$). However, our structure is slightly different since while the CO bond vector is nearly perpendicular to the inter-fragment vector, the CO fragment is tipped very slightly ($\theta = 87.89°$) such that the C-atom is closer to the $O_2$ fragment, while the opposite appears to be the case for the cross structure reported by Grein. The structure and geometric parameters are given in Fig. 6. Fig. 8 plots a relaxed scan of the torsional coordinate which connects the $o$-in global minimum with the cross structure. The barrier going from the $o$-in to the cross structure is only 4.0 cm$^{-1}$, while the barrier in the other direction is 1.5 cm$^{-1}$. The delocalized wells and small barriers shown in Figs. 7 and 8 make it essential to perform rovibrational calculations using a dense global grid since even the zero point vibration will cover large regions of the PES.

**Variational calculation of rovibrational levels**

The rovibrational levels of CO-$O_2$ were calculated using a variational method called DSL[56,57] which uses a product basis with discrete variable representation (DVR) functions (D)[58] for the stretches and spherical harmonic type functions (S) for the bends and a symmetry adapted Lanczos eigensolver (L). Each basis function is

$$f_{\alpha 0}(r_0) u^{JMP}_{l_1 l_2 m_2 K*}(\theta_1, \theta_2, \phi_2; \alpha, \beta, \gamma) \qquad (2)$$

where $f_{\alpha 0}(r_0)$ is a DVR function, $u^{JMP}_{l_1 l_2 m_2 K*}$ is a parity adapted rovibrational function and $\alpha$, $\beta$, and $\gamma$ are Euler angles. $P = 0$ and 1 correspond to even and odd parity, respectively. $(-1)^{J+P} = 1$ and -1 correspond to spectroscopic $e$ and $f$ parity, respectively. The vibrational coordinates are the polyspherical coordinates ($r_1, r_2, r_0, \theta_1, \theta_2, \varphi_2$) associated with the vector $r_1$ (for CO), the vector $r_2$ (for $O_2$), and the Jacobi vector $r_0$. Because the intramonomer vibrational frequencies are much higher than the intermonomer frequencies, it is justified to fix $r_1$ and $r_2$ to their respective ground state values. $J$ and $K*$ are labels for the angular momentum operator and its projection on the body-fixed z-axis. In the DSL method, potential energy integrals are computed with Gauss quadrature. Eigenvalues and eigenvectors are determined with the symmetry-adapted Lanczos (SAL) algorithm.[59,60] A thorough description of the DSL method applied to a similar system, $(N_2O)_2$, is reported in Refs. 46, 51, and the method was also recently applied to CO-$N_2$.[61] The calculation is carried out with the RV4 code[62] that implements the DSL method.

The rovibrational levels we report are computed with an angular basis having $l_{max} = m_{max} = 37$ (the same $l_{max}$ for $l_1$ and $l_2$) together with 38 Gauss-Legendre quadrature points for $\theta_1$ and $\theta_2$, and 80 equally spaced trapezoid points in the range $[0, 2\pi]$ for $\varphi_2$, with the first point zero. This bend basis is probably larger than necessary. For $r_0$ we use the efficient tridiagonal Morse DVR basis with the same parameters as used for CO-N$_2$. Convergence errors for levels calculated with this basis are estimated to be smaller than 0.001 cm$^{-1}$. The rotational constants for CO and O$_2$ are taken to be the experimental ground state values of 1.9225125 cm$^{-1}$ and 1.437678 cm$^{-1}$, respectively.[63,64] The masses are m(C) = 12.000 u, and m(O) = 15.9949146221 u.

In our calculation, we use the full permutation-inversion group of CO-O$_2$, G$_4$, consisting of four symmetries, $A^+$, $B^+$, $A^-$, and $B^-$, where $A/B$ labels the symmetric/anti-symmetric of the two O atoms of O$_2$, and +/- labels even/odd parity. By nuclear spin symmetry, only $B$ levels are allowed, corresponding to odd values of $n$(O$_2$). We assigned approximate quantum labels, $n$(O$_2$), $j$(CO), and $K^*$ to these calculated levels, where $K^*$ is the sum of the projections of the monomer angular momenta $n$(O$_2$) and $j$(CO) on the intermolecular axis. The $n$(O$_2$) and $j$(CO) labels were assigned using free-rotor energies, while $K^*$ was assigned by analyzing the wavefunctions.[65] The levels were thus organized into $K^*$-stacks which were then fitted with the empirical energy expression of Eq. 1. The resulting stack origins and $B$-values are given in Table 5, where each stack is given a label in order of increasing energy: A, B, C, etc. Energies are given relative to the origin of the first stack, A, which itself lies 1.757 cm$^{-1}$ above the hypothetical ground state with $n$(O$_2$) = 0. The calculated levels themselves are given in Table A-3 of the ESI for $J$ = 0 to 5, together with assigned $K^*$-values and stack labels.

To further characterize the stacks we extracted expectation values of $M_{n(O2)}$ and $M_{j(CO)}$, the projections of $n$(O$_2$) and $j$(CO), respectively, on the intermolecular axis. This calculation is straightforward since $M_{n(O2)}$ and $M_{j(CO)}$ are basis function labels (in the notation of Eq. 2, they correspond to $m_2$ and $m_1$; note that $m_1 = K^* - m_2$ is omitted in Eq. 2 because it is not an independent index). It turns out that the calculated $M_{n(O2)}$ and $M_{j(CO)}$ values, shown in the last column of Table 5, are similar for all the levels in a stack, helping to confirm the stack assignments. The values are all close to integers except for stacks F and G which are a mixture of $(M_{n(O2)}, M_{j(CO)})$ = (0, 1) and (1, 0) states.

The current calculation does *not* take into account the electronic spin term. Nevertheless, the calculation actually reproduces some aspects of the observed spectrum quite well, and helps to explain the observed energy level patterns. As shown in the following section, we can establish a convincing correspondence between experiment and theory and assign quantum labels to each observed $K$-stack.

## Beyond the free rotor picture: comparison of experiment and theory

A useful precedent for CO-O$_2$ is Ar-O$_2$, for which a molecular beam magnetic resonance spectrum was observed,[15] and calculations were carried out,[16,17] in the 1980s. These calculations indicated that the rotation of O$_2$ was hindered, but still relevant (so $n$(O$_2$) was still useful), but that the electron spin readily decoupled from the O$_2$ rotation (so $j$(O$_2$) was not so useful). The useful quantum numbers were the projections of $S$ and $n$(O$_2$) on the intermolecular axis, $M_S$ and $M_{n(O2)}$. A qualitative calculated result for Ar-O$_2$ is shown in Fig. 1 of Ref. 16. There are four low lying Ar-$O_2$ eigenstates, all with $S = 1$ (of course) and $n(O_2) = 1$. Lowest in energy is a $K = 0$ stack with $|M_S, M_{n(O2)}\rangle = 2^{-\frac{1}{2}} (|1, -1\rangle + |-1, 1\rangle)$; here $M_S$ and $M_{n(O2)}$ are anti-aligned, leaving zero net projection on the intermolecular axis. Next in energy was a $K = 2$ stack, with $|M_S, M_{n(O2)}\rangle = 2^{-\frac{1}{2}} (|1, 1\rangle \pm |-1, -1\rangle)$, followed by a $K = 1$ stack with $|M_S, M_{n(O2)}\rangle = 2^{-\frac{1}{2}} (|0, 1\rangle \pm |0, -1\rangle)$, and finally another $K = 0$ stack with $|M_S, M_{n(O2)}\rangle = 2^{-\frac{1}{2}} (|1, -1\rangle - |-1, 1\rangle)$. Further states with $n(O_2) = 1$, $M_{n(O2)} = 0$ were shifted to higher energy by the anisotropy of the Ar-$O_2$ potential.

This theoretical Ar-$O_2$ result[16] agrees with our observations for the lower states of CO-$O_2$. Specifically, the first two Ar-$O_2$ stacks, with $K = 0$ and 2, correspond to our observed $K = 0$ stack in group 1 and our observed $K = 2$ stack in group 2. Continuing upward, however, our other stacks have no Ar-$O_2$ analogs since they correlate with $j(CO) = 1$ and 2. The Ar-$O_2$ result suggests that CO-$O_2$ should have another $K = 1$ stack starting at roughly 4 cm$^{-1}$, which is what we also expect from the free rotor level $(n(O_2), j(O_2), j(CO)) = (1, 1, 0)$ in Fig. 5. This $K = 1$ stack would be the lowest stack of "group 3", but has not been assigned, presumably due to its higher energy and resulting low population.

We know from experiment and theory that $K$ is a 'good' quantum label, and we found from theory that the individual projections, $M_{n(O2)}$ and $M_{n(CO)}$, are also characteristic for the calculated stacks (Table 5). And finally we know from Ar-$O_2$ that the spin, $S$, readily uncouples from $j(O_2)$ when $O_2$ rotation is hindered, leaving $n(O_2)$, $M_S$, and $M_{n(O2)}$ as meaningful labels, rather than $j(O_2)$ and $M_{j(O2)}$. It therefore seems appropriate to use these good $M$-projection labels in order to go beyond the free rotor picture described above (Fig. 5). This is done in Table 6, which compares experiment and theory, revealing the correspondence between observed and calculated $K$-stacks. Here each observed stack has been labelled with the help of the *ab initio* results (Table 5), using $n(O_2)$ and $j(CO)$ together with $M_S$, $M_{n(O2)}$, and $M_{j(CO)}$. Note that the $K$-value of each stack is equal to the absolute sum of $M_S$, $M_{n(O2)}$, and $M_{j(CO)}$, as expected. The calculated (no-spin) $K$-values, $K^*$ from Table 5, do not include $M_S$, so they are not equal to the observed $K$-values. Instead, $K = K^* - 1$ for group 1, and $K = K^* + 1$ for group 2. Note also that $M_S$ and $M_{n(O2)}$ are aligned for group 2, and (mostly) anti-aligned for group 1, as expected.

There is rather striking agreement between the observed and calculated stack origins in Table 6, which gives us confidence that we are indeed on the right track in labelling the energy levels. $B$-values are not shown in Table 6, but are available from Tables 3 – 5. The ranges of the observed and calculated $B$-values are roughly similar, mostly falling between about 0.072 and 0.080 cm$^{-1}$, but the agreement in detail between experiment and theory is only limited. This is not surprising since electron spin, neglected so far in the theory, is almost certain to have a significant effect on dimer rotation.

Note that theoretical $K^* = 0$ stacks B/C should produce a group 1 stack with $K = 1$, but this has not yet been observed. Similarly, theoretical stacks E and F (Table 5) have no experimental counterpart so far.

## Conclusions

Our analysis accounts for most of the stronger observed lines in the spectrum of CO-$O_2$ and many weaker ones as well. However, there are still some unassigned features, which is not surprising considering the complexity of the CO-$O_2$ energy level scheme. There is considerable unassigned structure in the region from about 2140 to 2143 cm$^{-1}$ which becomes more prominent as the fraction of $O_2$ in the expansion gas mix is increased. Some of this structure may be due to CO-$O_2$, but based on the line density and concentration dependence we think that some may also be due to larger clusters such as CO-$(O_2)_2$. In the region of Fig. 1, there are notable unassigned lines at 2145.803, 2145.813, 2145.936, 2145.991, and 2146.047 cm$^{-1}$, and in the corresponding mirror-image region there are lines at 2139.615, 2139.612, 2139.591, 2139.593, 2139.545, 2139.534, and 2139.469 cm$^{-1}$. In the region of the $K = 2 \leftarrow 1$ band of group 1 and the $K = 4 \leftarrow 3$ band of group 2, there are lines at 2149.841 and 2149.996 cm$^{-1}$. It is plausible to suppose that some of these unassigned lines could belong to the as yet unassigned "group 3", correlating with $(n(O_2), j(O_2)) = (1, 1)$. More specifically, we anticipate that group 3 could give a $K = 2 \leftarrow 1$ band in the 2145 region, a $K = 1 \leftarrow 2$ band in the 2139 region, and a $K = 3 \leftarrow 2$ band in the 2150 region.

By analogy with CO-$N_2$, it should be possible to observe extensive pure rotational spectra of CO-$O_2$, thereby extending and refining the current results. In the microwave region, the spectrum will depend on a very small induced dipole moment, but this weakness can be compensated by the high sensitivity of the Fourier transform microwave technique.[29] Stronger transitions depending on the permanent dipole moment of CO are expected in the millimeter-wave region. Predicted transition frequencies are easily calculated from our experimental energy levels in Tables 1 and 2. For example the strongest $K = 0$ group 1 microwave series should fall approximately at 4626, 9250, 13823, 18360 MHz, and a $K = 1 \leftarrow 0$ millimeter $Q$-branch should fall approximately at 82355, 82796, 83454, 84264 MHz.

In conclusion, detailed infrared spectra of the weakly-bound CO-$O_2$ complex have been observed in the CO fundamental band region ($\approx$2150 cm$^{-1}$) using a tunable quantum cascade laser source to probe a pulsed slit-jet supersonic expansion. The spectra were assigned in terms of a number of stacks of rotational levels having well-defined values of $K$, the projection of the total angular momentum on the intermolecular axis. These stacks were divided into two groups, with no observed transitions between the groups. The groups are believed to correspond to different projections of $S (= 1)$, the $O_2$ unpaired electron spin, and to correlate with the two lowest rotational levels of $O_2$, $(n(O_2), j(O_2)) = (1, 0)$ and $(1, 2)$. In the ground vibrational state ($v(CO) = 0$), there are two and three stacks assigned in the two groups, respectively. In the excited state ($v(CO) = 1$), there are five stacks assigned in each group. The relative energies of the two groups are not determined precisely, but from intensities it appear that the $(n(O_2), j(O_2)) = (1, 2)$ group lies about 2 cm$^{-1}$ above the (1, 0) group. The *ab initio* calculations reported here provide a qualitative explanation of the experimental rotational stacks and enable the assignment of detailed quantum labels to the stacks, even though the calculations do not so far include spin. A better understanding, and extension of the experimental results to further energy levels, should be possible by including the effects of electron spin.

**Acknowledgements**

The financial support of the Canadian Space Agency and the Natural Sciences and Engineering Research Council of Canada is gratefully acknowledged. RD is supported by the US National Science Foundation (No. CHE-1566246). We thank K. H. Michaelian for the loan of the quantum cascade laser system.

Table 1. Experimental CO-O$_2$ energy levels of group 1, correlating with $(n(O_2), j(O_2)) = (1, 0)$ (in cm$^{-1}$).[a]

| J | v(CO) = 0  K = 0e | v(CO) = 0  K = 1e | v(CO) = 0  K = 1f | v(CO) = 1  K = 0e | v(CO) = 1  K = 1e | v(CO) = 1  K = 1f | v(CO) = 1  K = 2e | v(CO) = 1  K = 2f | v(CO) = 1  K = 0'e | v(CO) = 1  K = 0''e |
|---|---|---|---|---|---|---|---|---|---|---|
| 0 | 0.0000 | | | 0.0000 | | | | | 9.1293 | 10.1004 |
| 1 | 0.1543 | 2.8945 | 2.9014 | 0.1548 | 2.8628 | 2.8706 | | | 9.2691 | 10.1762 |
| 2 | 0.4628 | 3.2041 | 3.2246 | 0.4633 | 3.1716 | 3.1955 | 10.0276 | 10.0271 | 9.5472 | 10.3705 |
| 3 | 0.9239 | 3.6643 | 3.7076 | 0.9248 | 3.6310 | 3.6804 | 10.4789 | 10.4787 | 9.9620 | 10.7072 |
| 4 | 1.5364 | 4.2724 | 4.3471 | 1.5371 | 4.2379 | 4.3222 | 11.0746 | 11.0776 | 10.5108 | 11.1986 |
| 5 | 2.2976 | 5.0246 | 5.1401 | 2.2990 | 4.9891 | 5.1198 | 11.7908 | 11.8217 | 11.1915 | 11.8733 |
| 6 | 3.2058 | 5.9182 | 6.0818 | 3.2068 | 5.8816 | 6.0668 | 12.7011 | 12.7074 | 11.9840 | 12.6066 |
| 7 | 4.2579 | 6.9503 | 7.1678 | 4.2600 | 6.9135 | 7.1567 | | | | |
| 8 | | | | 5.4554 | 8.0825 | | | | | |

[a] These are 'experimental' energies, based on term value fits, except that it is necessary to fix by fitting one interval between ground state levels of opposite parity, specifically the 0.1543 cm$^{-1}$ interval between the two lowest levels. This interval can be experimentally determined in the future by observing pure rotational spectra of CO-O$_2$. All v(CO) = 1 energies are expressed relative to the origin value, 2142.6942 cm$^{-1}$. All v(CO) = 1 energies are expressed relative to the origin value, 2142.6942 cm$^{-1}$.

Table 2. Experimental CO-$O_2$ energy levels of group 2, correlating with $(n(O_2), j(O_2)) = (1, 2)$ (in cm$^{-1}$).[a]

| J | v(CO) = 0 K = 2e | v(CO) = 0 K = 2f | v(CO) = 0 K = 3e | v(CO) = 0 K = 3f | v(CO) = 0 K = 1e | v(CO) = 0 K = 1f |
|---|---|---|---|---|---|---|
| 1 | | | | | 2.9824 | 2.9858 |
| 2 | 0.1471 | 0.1462 | | | 3.2723 | 3.2837 |
| 3 | 0.5894 | 0.5851 | 3.2125 | 3.2120 | 3.7101 | 3.7388 |
| 4 | 1.1840 | 1.1714 | 3.8031 | 3.8012 | 4.2978 | 4.3346 |
| 5 | 1.9353 | 1.9085 | 4.5460 | 4.5429 | | |
| 6 | 2.8481 | 2.7917 | 5.4459 | 5.4346 | | |

| J | v(CO) = 1 K = 2e | v(CO) = 1 K = 2f | v(CO) = 1 K = 3e | v(CO) = 1 K = 3f | v(CO) = 1 K = 1e | v(CO) = 1 K = 1f | v(CO) = 1 K = 4e | v(CO) = 1 K = 4f | v(CO) = 1 K = 2'e | v(CO) = 1 K = 2'f |
|---|---|---|---|---|---|---|---|---|---|---|
| 1 | | | | | 2.9647 | 2.9682 | | | | |
| 2 | 0.1473 | 0.1466 | | | 3.2551 | 3.2673 | | | 9.9422 | 9.9366 |
| 3 | 0.5892 | 0.5847 | 3.1941 | 3.1938 | 3.6928 | 3.722 | | | 10.3936 | 10.3643 |
| 4 | 1.1845 | 1.1726 | 3.7861 | 3.7854 | 4.2796 | 4.3236 | 10.3675 | 10.367 | 11.0071 | 10.9325 |
| 5 | 1.9362 | 1.9084 | 4.5327 | 4.5287 | | | 11.0882 | 11.0863 | | |
| 6 | 2.8484 | 2.7936 | 5.4368 | 5.4258 | | | 11.9561 | 11.9579 | | |
| 7 | | | 6.4911 | 6.4839 | | | 12.9758 | 12.9684 | | |

[a] These are 'experimental' energies, based on term value fits. The zero of energy is simply the (calculated) origin value of the lowest $K = 2$ stack, and this origin lies above that of Table 1 by an unknown amount X which is approximately equal to 2 cm$^{-1}$ (see text). It is necessary to fix by fitting one interval between ground state levels of opposite parity, specifically the 0.4423 cm$^{-1}$ interval between the two lowest K = 2e levels. This interval can be experimentally determined in the future by observing pure rotational spectra of CO-$O_2$. All v(CO) = 1 energies are expressed relative to the origin value, 2142.6942 cm$^{-1}$.

Table 3. Effective parameters for observed CO-O$_2$ $K$-stacks of group 1 (in cm$^{-1}$). [a]

| | v(CO) = 0<br>K = 0e | v(CO) = 0<br>K = 1e,f | v(CO) = 1<br>K = 0e | v(CO) = 1<br>K = 1e,f | v(CO) = 1<br>K = 2e,f | v(CO) = 1<br>K = 0'e | v(CO) = 1<br>K = 0"e |
|---|---|---|---|---|---|---|---|
| σ | 0.0 | 2.8196 | 2142.6942 | 2145.4824 | 2152.5710 | 2151.8235 | 2152.7946 |
| σ rel | | | 0.0 | 2.7882 | 9.8768 | 9.1293 | 10.1004 |
| B | 0.07724 | 0.07911 | 0.07729 | 0.07921 | 0.07540 | 0.07015 | 0.04203 |
| b | | -0.00370 | | -0.00425 | | | |
| 10$^5$×D | 2.2 | 3.7 | 2.2 | 3.6 | 2.6 | 5.1 | -65. |
| 10$^5$×d | | -0.4 | | -0.2 | -0.4 | | |

[a] The $J$ = 5 level of $K$ = 2e was omitted from the fit due to perturbation. The $K$ = 0" stack was fitted only for $J$ = 1 – 4, with its origin fixed at $J$ = 0. σ rel is the origin relative to the lowest origin for v(CO) = 1. Note that the σ rel values for v(CO) = 1 are quite similar to σ for v(CO) = 0.

Table 4. Effective parameters for observed CO-O$_2$ $K$-stacks of group 2 (in cm$^{-1}$).[a]

| $J$ | $v(CO) = 0$ $K = 2e,f$ | $v(CO) = 0$ $K = 3e,f$ | $v(CO) = 0$ $K = 1e,f$ | $v(CO) = 1$ $K = 2e,f$ | $v(CO) = 1$ $K = 3e,f$ | $v(CO) = 1$ $K = 1e,f$ | $v(CO) = 1$ $K = 4e,f$ | $v(CO) = 1$ $K = 2'e,f$ |
|---|---|---|---|---|---|---|---|---|
| $\sigma$ | 0.0 | 2.9917 | 2.9096 | 2142.6943 | 2145.6663 | 2145.5861 | 2152.7732 | 2152.4879 |
| $\sigma$ rel | | | | 0.0 | 2.9720 | 2.8918 | 10.0789 | 9.7936 |
| $B$ | 0.07322 | 0.07347 | 0.07394 | 0.07323 | 0.07379 | 0.07411 | 0.07196 | 0.07281 |
| $b$ | | | -0.00255 | | | -0.00264 | | |
| $10^5 \times D$ | -2.6 | -2.2 | -0.6 | -2.7 | -2.2 | 0.3 | -0.9 | -4.4 |
| $10^5 \times d$ | 3.1 | | 3.3 | 3.1 | | 3.5 | | 18.8 |
| $10^7 \times h$ | | 1.5 | | | 0.6 | | | |

[a] Relative to group 1 (Table 3), the group 2 origins are higher by an unknown amount which is approximately equal to 2 cm$^{-1}$ (see text). $\sigma$ rel is the origin relative to the lowest origin for $v(CO) = 1$.

Table 5. Fit parameters and assigned quantum numbers for theoretical (no spin) $K$-stacks of CO-O$_2$.

| stack label | $K^*$ | origin | $B$ | assigned $(n_{O2}, j_{CO})$ | assigned $|Mn_{(O2)}, Mj_{(CO)}\rangle$ | expectation value $(Mn_{(O2)}, Mj_{(CO)})$ |
|---|---|---|---|---|---|---|
| A | 1 $e,f$ | 0.000 | 0.0791 | (1, 0) | $2^{-\frac{1}{2}}(|1, 0\rangle \pm |-1, 0\rangle)$ | (0.99, 0.01) |
| B | 0 $f$ | 2.736 | 0.0790 | (1, 1) | $2^{-\frac{1}{2}}(|1, -1\rangle - |-1, 1\rangle)$ | (1.01, -1.01) |
| C | 0 $e$ | 2.877 | 0.0785 | (1, 1) | $2^{-\frac{1}{2}}(|1, -1\rangle + |-1, 1\rangle)$ | (0.91, -0.91) |
| D | 2 $e,f$ | 2.815 | 0.0777 | (1, 1) | $2^{-\frac{1}{2}}(|1, 1\rangle \pm |-1, -1\rangle)$ | (1.00, 0.99) |
| E | 0 $e$ | 6.560 | 0.0730 | (1, 0) | $|0, 0\rangle$ | (0.11, -0.11) |
| F | 1 $e,f$ | 7.833 | 0.0762 | (1, 1) | $2^{-\frac{1}{2}}(|1, 0\rangle \pm |-1, 0\rangle)$, $2^{-\frac{1}{2}}(|0, 1\rangle \pm |0, -1\rangle)$ | (0.41, 0.59) |
| G | 1 $e f$ | 9.808 | 0.0733 | (1, 1) | $2^{-\frac{1}{2}}(|0, 1\rangle \pm |0, -1\rangle)$, $2^{-\frac{1}{2}}(|1, 0\rangle \pm |-1, 0\rangle)$ | (0.50, 0.50) |
| H | 3 $e,f$ | 9.557 | 0.0770 | (1, 2) | $2^{-\frac{1}{2}}(|1, 2\rangle \pm |-1, -2\rangle)$ | (1.02, 1.96) |
| I | 1 $e,f$ | 10.056 | 0.0772 | (1, 2) | $2^{-\frac{1}{2}}(|1, -2\rangle \pm |-1, 2\rangle)$ | (-0.80, 1.80) |
| J | 1 $e,f$ | 12.397 | 0.0766 | (3, 0) | $2^{-\frac{1}{2}}(|3, 0\rangle \pm |-3, 0\rangle)$ | (2.90, 0.10) |
| K | 3 $e,f$ | 13.365 | 0.0705 | (1, 2) | $2^{-\frac{1}{2}}(|1, 1\rangle \pm |-1, -1\rangle)$ | (1.00, 1.00) |

[a] Origins and $B$-values are in cm$^{-1}$, with origins relative to that of the first stack. Note that $K^* = M_{n(O2)} + M_{j(CO)}$. Note also that stacks F and G are highly mixed. Since each basis function (Eq. 2) has a well-defined spectroscopic parity $e/f$ label, by construction the $e/f$ label is associated with the +/- combination, respectively, allowing assignment of $e/f$ labels to the split $K^* = 0$ stack B/C (see e.g. Eq. 9 of Ref. 59).

Table 6. Theoretical *ab initio* (no-spin) $K$-stack origins and labels of CO-$O_2$, together with the observed stacks and their proposed labels in terms of angular momentum projections ($M_S$, $M_{n(O2)}$, $M_{j(CO)}$) on the intermolecular axis. [a]

| | Theory, no spin[b] | | Experiment, group 1 $M_S$ anti-aligned with $Mn_{(O2)}$ $(n(O_2), j(O_2)) = (1, 0)$ | | | Experiment, group 2 $M_S$ aligned with $Mn_{(O2)}$ $(n(O_2), j(O_2)) = (1, 2)$ | | |
|---|---|---|---|---|---|---|---|---|
| stack label | origin | $(j_{CO}, K)$ | $\| M_S, M_{n(O2)}, M_{j(CO)} \rangle$ | origin | $(j_{CO}, K)$ | $\| M_S, M_{n(O2)}, M_{j(CO)} \rangle$ | origin |
| A | 0 | (0, 0) | $2^{-½} (\|-1, 1, 0\rangle + \|1, -1, 0\rangle)$ | 0.00 | (0, 2) | $2^{-½} (\|1, 1, 0\rangle \pm \|-1, -1, 0\rangle)$ | 0.00 |
| B/C | 2.74 / 2.88 | | | | (1, 1) | $2^{-½} (\|1, 1, -1\rangle \pm \|-1, -1, 1\rangle)$ | 2.89 |
| D | 2.82 | (1, 1) | $2^{-½} (\|-1, 1, 1\rangle \pm \|1, -1, -1\rangle)$ | 2.79 | (1, 3) | $2^{-½} (\|1, 1, 1\rangle \pm \|-1, -1, -1\rangle)$ | 2.97 |
| G | 9.82 | (1, 0') | $2^{-½} (\|-1, 1, 0\rangle + \|1, -1, 0\rangle)$ | 9.13 | (1, 2') | $2^{-½} (\|1, 1, 0\rangle \pm \|-1, -1, 0\rangle)$ | 9.79 |
| H | 9.56 | (2, 2) | $2^{-½} (\|-1, 1, 2\rangle \pm \|1, -1, -2\rangle)$ | 9.88 | (2, 4) | $2^{-½} (\|1, 1, 2\rangle \pm \|-1, -1, -2\rangle)$ | 10.08 |
| I | 10.06 | (1, 0") | $2^{-½} (\|-1, 0, 1\rangle + \|1, 0, -1\rangle)$ | 10.10 | | | |

[a] Origins in cm$^{-1}$. Experimental origins are for v(CO) = 1, which is more complete, but v(CO) = 0 is similar. Origins of experimental group 1 are relative to 2142.6942 cm$^{-1}$, the origin of the first stack. Origins of experimental group 2 are relative to the origin of its first stack at 2142.6943 cm$^{-1}$. Group 2 is thought to be about 2 cm$^{-1}$ higher than group 1.

[b] The theoretical $n(O_2)$, $j(CO)$, $M_{n(O2)}$, and $M_{j(CO)}$ assignments for each stack are given in Table 5, and agree with the experimental values given here. Since electron spin is not included in the theory, the theoretical $K^*$ and experimental $K$-values differ. For group 1, $K = K^* - 1$, and for group 2, $K = K^* + 1$, where $K^*$ is the theoretical value from Table 5.

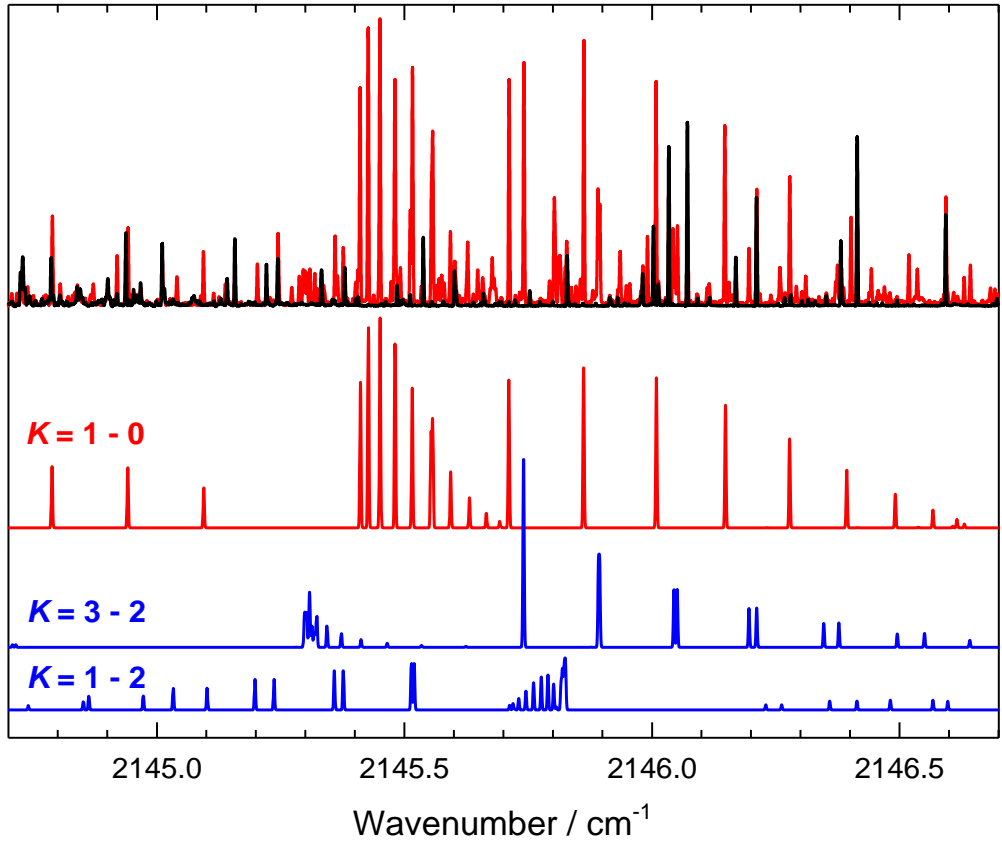

Fig. 1. Part of the observed (top trace) and simulated spectrum of CO-$O_2$. The top red trace (CO + $O_2$ + He gas mix) contains CO-$O_2$ and $(CO)_2$ transitions, while the top black trace (CO + He mix), plotted in front, contains only $(CO)_2$ transitions. The simulated spectra (lower 3 traces) assume an effective temperature of 2.2 K.

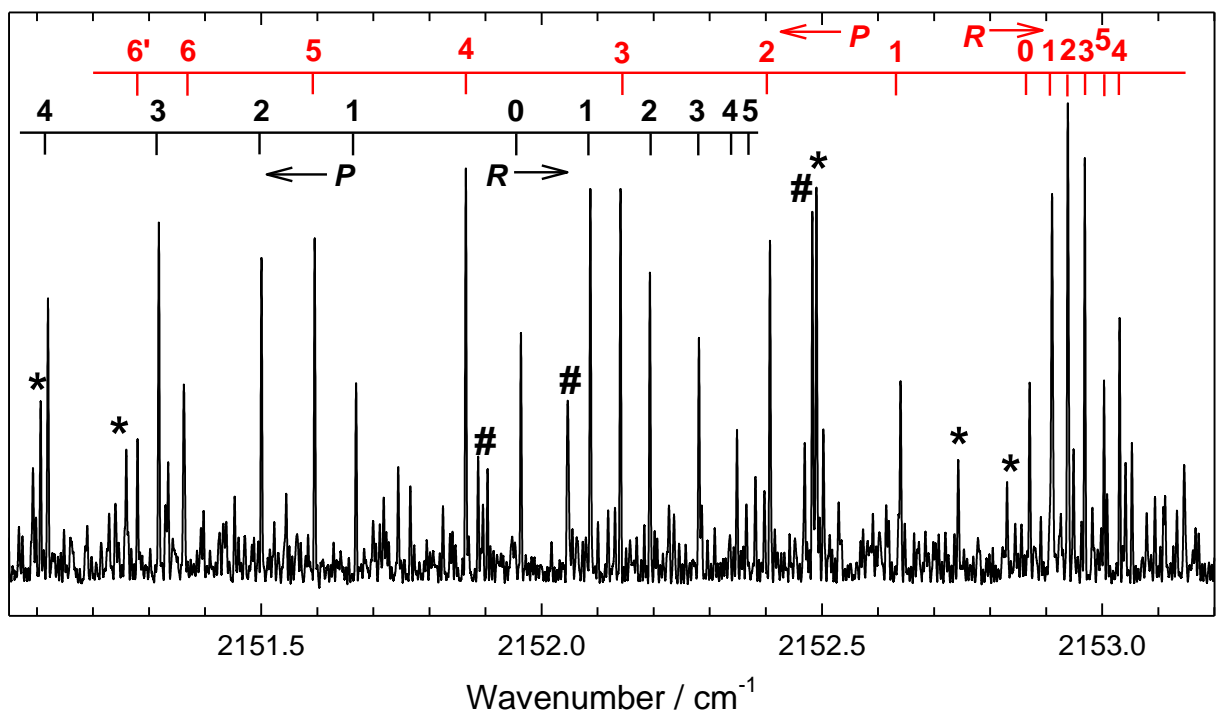

Fig. 2. Observed spectrum of CO-O$_2$ showing the $K = 0' \leftarrow 0$ and $0'' \leftarrow 0$ bands, with the latter labelled in red (the line labelled $P(6')$ arises from the level crossing discussed in the text). Asterisks indicate (CO)$_2$ transitions, and # indicates transitions of the $K = 2' \leftarrow 2$ band of group 2.

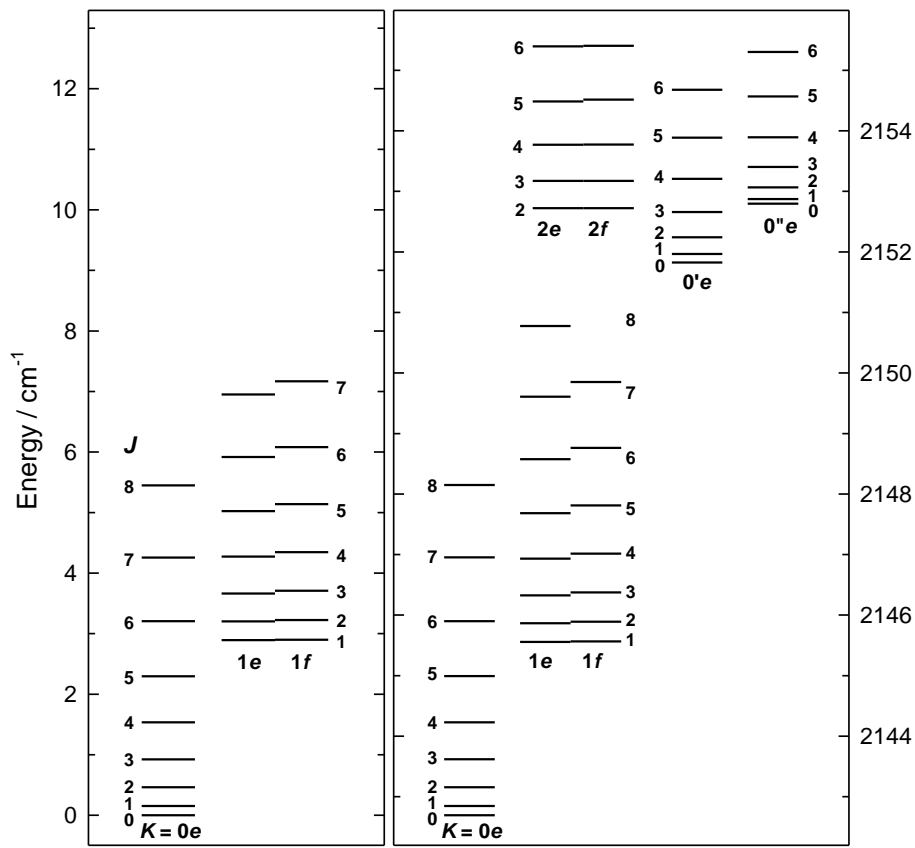

Fig. 3. Experimental CO-O$_2$ energy levels belonging to group 1, with ground state (v(CO) = 0) on the left and the excited state (v(CO) = 1) on the right. The levels belong to stacks with well-defined *K*-values.

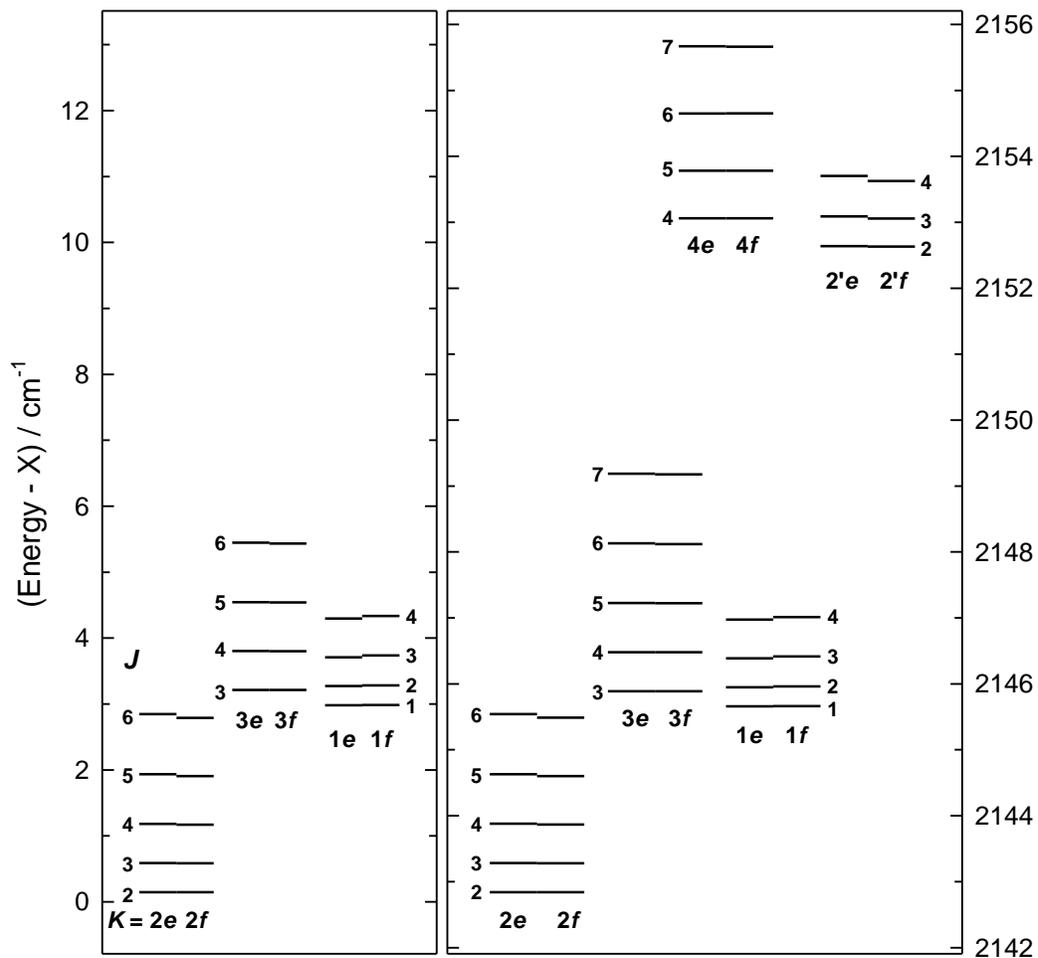

Fig. 4. Experimental CO-O$_2$ energy levels belonging to group 2, with ground state (v(CO) = 0) on the left and the excited state (v(CO) = 1) on the right. The energy of these levels relative to those in Fig. 3 is not known exactly because no observed transitions connect them. But we estimate from observed intensities that the quantity X is approximately 2 cm$^{-1}$.

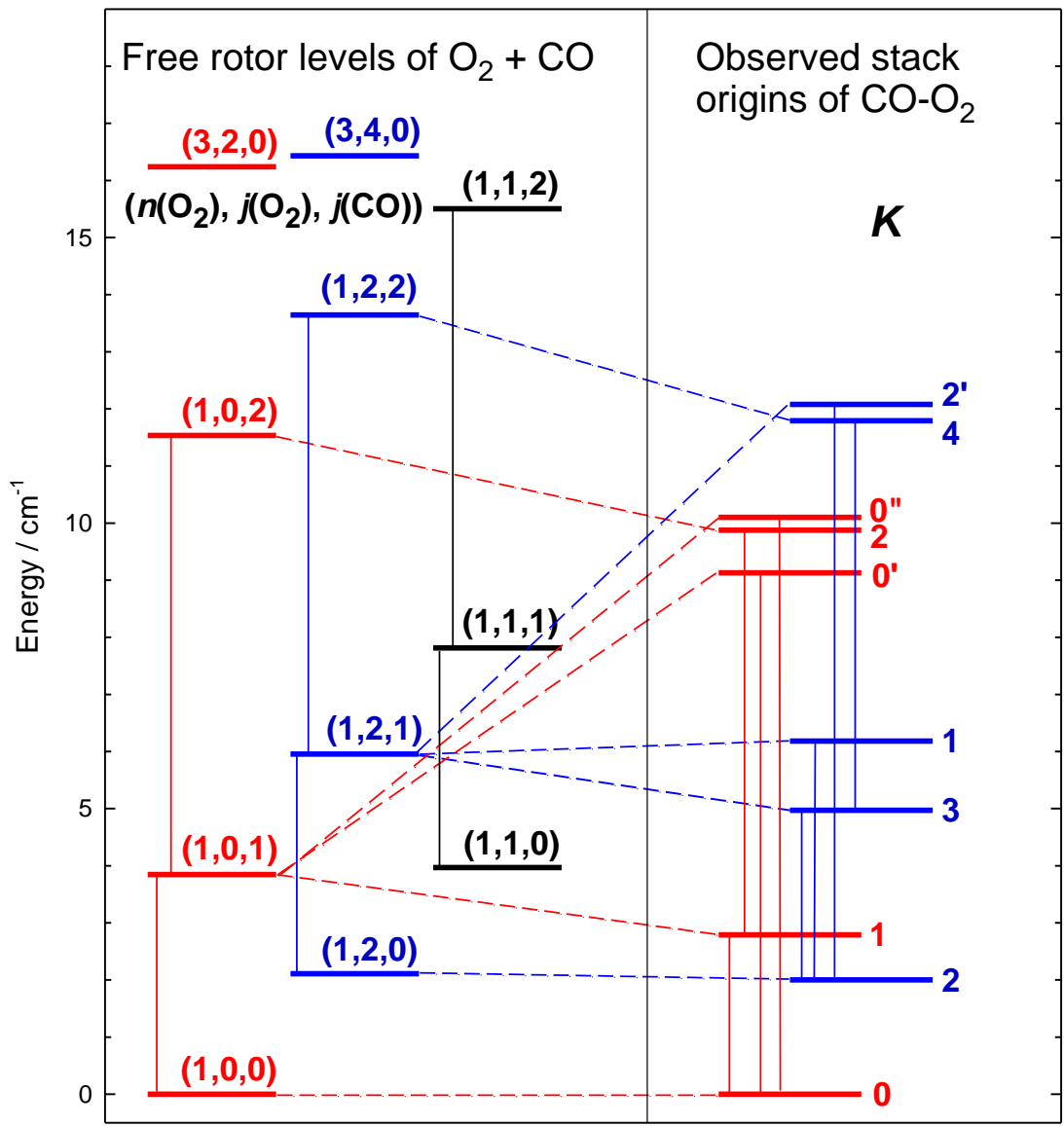

Fig. 5. Free rotor picture for CO + $O_2$ (left hand side) and observed stack origins for CO-$O_2$ (right hand side). Group 1 is red and group 2 blue. Dashed lines show proposed correlations. Thin vertical lines show allowed free-rotor transitions ($\Delta j$(CO) = ±1) on the left, and observed CO-$O_2$ bands on the right.

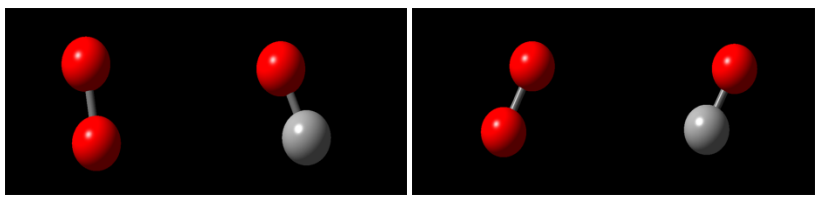

Fig. 6. The planar global minimum $o$-in structure (E = -119.3 cm$^{-1}$). Local minima are at $c$-in planar (E = -112.9 cm$^{-1}$) and cross-shaped (E = -116.8 cm$^{-1}$). The geometric parameters ($R$, $\theta_1$, $\theta_2$, $\varphi$) in Angstroms and degrees are (3.460 Å, 100.68°, 100.35°, 0°), (3.819 Å, 61.60°, 56.29°, 0°), (3.451 Å, 87.89°, 90°, 90°) for $o$-in, $c$-in, and $cross$ respectively.

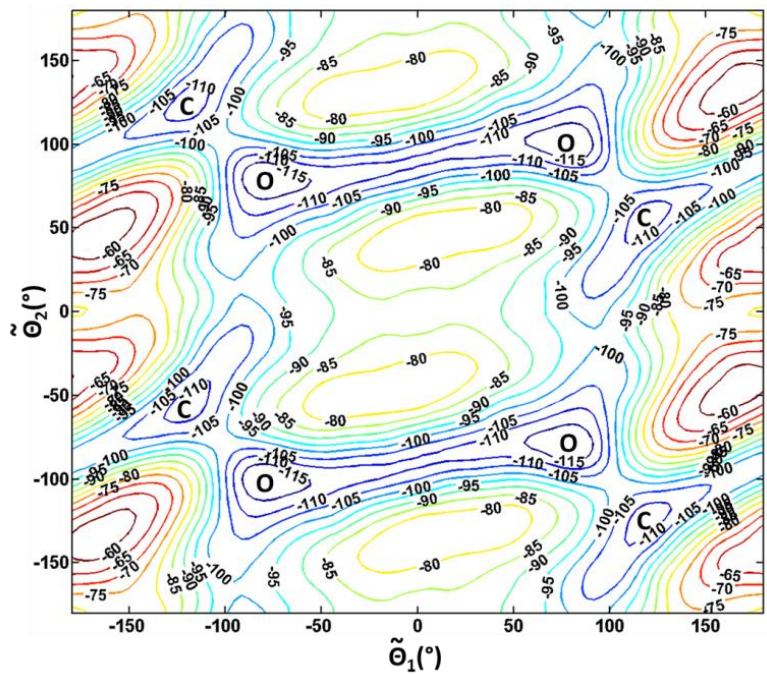

Fig. 7. Extended angles plot for planar geometries (angles in degrees and energies in cm$^{-1}$). For each pair of angles, the energy is minimized by varying the center-of-mass distance $R$. The wells corresponding to the $o$-in (global minimum) and $c$-in (local minimum) structures are labeled (**O** and **C** respectively). Fig. 8 shows the out-of-plane torsional path connecting $o$-in to the $cross$ structure.

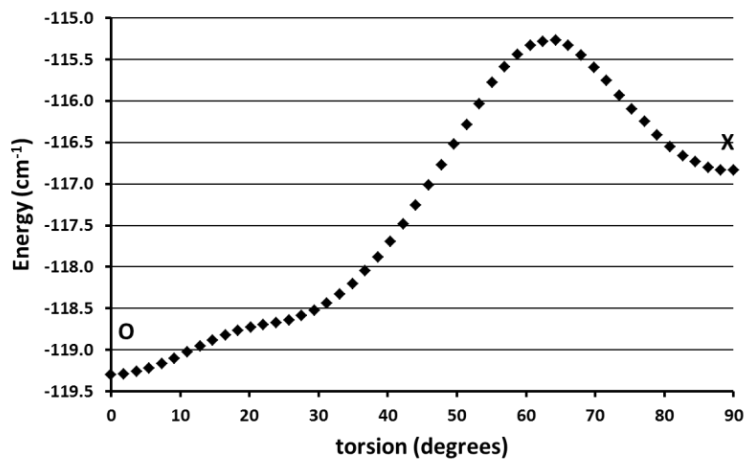

Fig. 8. Relaxed scan of torsional coordinate $\varphi$ connecting global *o*-in minimum with local cross-shaped minimum (denoted *X*). Images and structural parameters are given in Fig. 6.

Appendix to:
# Infrared spectrum and intermolecular potential energy surface of the CO – O₂ dimer

A.J. Barclay, A.R.W. McKellar, N. Moazzen-Ahmadi, Richard Dawes, Xiao-Gang Wang, and Tucker Carrington Jr.

Table A-1. Observed transitions of CO-O$_2$ in group 1, correlating with ($n$(O$_2$) = 1, $j$(O$_2$) = 0) (units of cm$^{-1}$). Note: The "calculated" positions correspond to the "experimental" energy levels from Table 1 of the paper.

| Upper $J, K\ e/f$ | Lower $J, K\ e/f$ | Observed | Calculated | Obs - Calc |
|---|---|---|---|---|
| 1,1f | 1,0 | 2145.411 | 2145.411 | 0.0000 |
| 2,1f | 2,0 | 2145.427 | 2145.427 | 0.0000 |
| 3,1f | 3,0 | 2145.451 | 2145.451 | 0.0000 |
| 4,1f | 4,0 | 2145.480 | 2145.480 | 0.0000 |
| 5,1f | 5,0 | 2145.516 | 2145.516 | 0.0000 |
| 6,1f | 6,0 | 2145.555 | 2145.555 | 0.0000 |
| 7,1f | 7,0 | 2145.593 | 2145.593 | 0.0000 |
| 8,1f | 8,0 | 2145.628 | 2145.628 | 0.0000 |
| 1,1e | 0,0 | 2145.557 | 2145.557 | -0.0001 |
| 2,1e | 1,0 | 2145.712 | 2145.712 | 0.0000 |
| 3,1e | 2,0 | 2145.862 | 2145.862 | 0.0000 |
| 4,1e | 3,0 | 2146.008 | 2146.008 | -0.0001 |
| 5,1e | 4,0 | 2146.147 | 2146.147 | -0.0001 |
| 6,1e | 5,0 | 2146.278 | 2146.278 | 0.0000 |
| 7,1e | 6,0 | 2146.402 | 2146.402 | 0.0000 |
| 8,1e | 7,0 | 2146.519 | 2146.519 | 0.0000 |
| 1,1e | 2,0 | 2145.094 | 2145.094 | 0.0001 |
| 2,1e | 3,0 | 2144.942 | 2144.942 | 0.0000 |
| 3,1e | 4,0 | 2144.789 | 2144.789 | 0.0000 |
| 4,1e | 5,0 | 2144.635 | 2144.635 | 0.0001 |
| 5,1e | 6,0 | 2144.478 | 2144.478 | 0.0001 |
| 6,1e | 7,0 | 2144.318 | 2144.318 | -0.0001 |
| 7,1e | 8,0 | 2144.157 | 2144.158 | -0.0003 |
| | | | | |
| 1,0 | 0,0 | 2142.849 | 2142.849 | 0.0001 |
| 2,0 | 1,0 | 2143.003 | 2143.003 | 0.0001 |
| 3,0 | 2,0 | 2143.156 | 2143.156 | 0.0000 |
| 4,0 | 3,0 | 2143.308 | 2143.307 | 0.0001 |
| 5,0 | 4,0 | 2143.457 | 2143.457 | 0.0000 |
| 6,0 | 5,0 | 2143.603 | 2143.603 | 0.0000 |
| 7,0 | 6,0 | 2143.748 | 2143.748 | 0.0000 |
| 8,0 | 8,1f | 2139.758 | 2139.758 | 0.0000 |
| 0,0 | 1,0 | 2142.540 | 2142.540 | -0.0001 |
| 1,0 | 2,0 | 2142.386 | 2142.386 | -0.0002 |
| 2,0 | 3,0 | 2142.234 | 2142.234 | 0.0000 |
| 3,0 | 4,0 | 2142.083 | 2142.083 | 0.0002 |
| 4,0 | 5,0 | 2141.934 | 2141.934 | -0.0001 |
| 5,0 | 6,0 | 2141.787 | 2141.787 | -0.0001 |
| 6,0 | 7,0 | 2141.643 | 2141.643 | 0.0000 |
| | | | | |
| 1,0 | 1,1f | 2139.948 | 2139.948 | 0.0000 |
| 2,0 | 2,1f | 2139.933 | 2139.933 | 0.0000 |
| 3,0 | 3,1f | 2139.911 | 2139.911 | 0.0000 |
| 4,0 | 4,1f | 2139.884 | 2139.884 | 0.0000 |
| 5,0 | 5,1f | 2139.853 | 2139.853 | 0.0000 |
| 6,0 | 6,1f | 2139.819 | 2139.819 | 0.0000 |
| 7,0 | 7,1f | 2139.787 | 2139.787 | 0.0000 |
| 8,0 | 8,1f | 2139.758 | 2139.758 | 0.0000 |
| 0,0 | 1,1e | 2139.800 | 2139.800 | 0.0001 |
| 1,0 | 2,1e | 2139.645 | 2139.645 | 0.0001 |
| 2,0 | 3,1e | 2139.493 | 2139.493 | 0.0000 |
| 3,0 | 4,1e | 2139.347 | 2139.347 | -0.0001 |
| 4,0 | 5,1e | 2139.207 | 2139.207 | 0.0000 |
| 5,0 | 6,1e | 2139.075 | 2139.075 | 0.0000 |
| 6,0 | 7,1e | 2138.951 | 2138.951 | 0.0000 |
| 2,0 | 1,1e | 2140.263 | 2140.263 | -0.0001 |
| 3,0 | 2,1e | 2140.415 | 2140.415 | -0.0001 |
| 4,0 | 3,1e | 2140.567 | 2140.567 | 0.0000 |
| 5,0 | 4,1e | 2140.721 | 2140.721 | 0.0001 |
| 6,0 | 5,1e | 2140.876 | 2140.876 | 0.0000 |
| 7,0 | 6,1e | 2141.036 | 2141.036 | 0.0000 |
| | | | | |
| 1,0' | 0,0 | 2151.963 | 2151.963 | 0.0000 |
| 2,0' | 1,0 | 2152.087 | 2152.087 | 0.0000 |
| 3,0' | 2,0 | 2152.193 | 2152.193 | 0.0000 |
| 4,0' | 3,0 | 2152.281 | 2152.281 | 0.0000 |
| 5,0' | 4,0 | 2152.349 | 2152.349 | -0.0008 |
| 6,0' | 5,0 | 2152.381 | 2152.381 | 0.0008 |
| 0,0' | 1,0 | 2151.669 | 2151.669 | 0.0000 |
| 1,0' | 2,0 | 2151.501 | 2151.501 | 0.0000 |
| 2,0' | 3,0 | 2151.318 | 2151.318 | 0.0000 |
| 3,0' | 4,0 | 2151.120 | 2151.120 | 0.0000 |
| 4,0' | 5,0 | 2150.907 | 2150.907 | 0.0000 |
| 5,0' | 6,0 | 2150.680 | 2150.680 | 0.0004 |
| 6,0' | 7,0 | 2150.420 | 2150.420 | -0.0006 |
| | | | | |
| 2,2e | 1,1e | 2149.827 | 2149.827 | -0.0005 |
| 3,2e | 2,1e | 2149.969 | 2149.969 | -0.0001 |
| 4,2e | 3,1e | 2150.104 | 2150.104 | -0.0002 |
| 5,2e | 4,1e | 2150.213 | 2150.213 | -0.0001 |
| 6,2e | 5,1e | 2150.367 | 2150.371 | -0.0035 |
| 2,2f | 1,1f | 2149.820 | 2149.820 | 0.0002 |
| 3,2f | 2,1f | 2149.948 | 2149.948 | 0.0002 |
| 4,2f | 3,1f | 2150.064 | 2150.064 | 0.0002 |
| 5,2f | 4,1f | 2150.169 | 2150.169 | 0.0002 |
| 6,2f | 5,1f | 2150.261 | 2150.262 | -0.0001 |
| 2,2e | 2,1f | 2149.498 | 2149.497 | 0.0005 |
| 3,2e | 3,1f | 2149.466 | 2149.466 | 0.0001 |
| 4,2e | 4,1f | 2149.422 | 2149.422 | 0.0002 |
| 5,2e | 5,1f | 2149.345 | 2149.345 | -0.0004 |
| 2,2f | 2,1e | 2149.517 | 2149.517 | -0.0002 |
| 3,2f | 3,1e | 2149.508 | 2149.509 | -0.0002 |
| 4,2f | 4,1e | 2149.499 | 2149.499 | -0.0002 |
| 5,2f | 5,1e | 2149.491 | 2149.491 | -0.0002 |
| 6,2f | 6,1e | 2149.484 | 2149.484 | 0.0001 |
| | | | | |
| 5,2e | 4,0 | 2152.949 | 2152.949 | 0.0002 |
| 6,2e | 5,0 | 2153.094 | 2153.098 | -0.0038 |
| 5,2e | 6,0 | 2151.280 | 2151.279 | 0.0003 |
| | | | | |
| 1,0" | 0,0 | 2152.871 | 2152.870 | 0.0001 |
| 2,0" | 1,0 | 2152.910 | 2152.910 | 0.0000 |
| 3,0" | 2,0 | 2152.939 | 2152.939 | 0.0000 |
| 4,0" | 3,0 | 2152.969 | 2152.969 | -0.0001 |
| 5,0" | 4,0 | 2153.031 | 2153.031 | -0.0002 |
| 6,0" | 5,0 | 2153.003 | 2153.003 | 0.0000 |
| 0,0" | 1,0 | 2152.640 | 2152.640 | 0.0000 |
| 1,0" | 2,0 | 2152.408 | 2152.408 | -0.0001 |
| 2,0" | 3,0 | 2152.141 | 2152.141 | 0.0000 |
| 3,0" | 4,0 | 2151.865 | 2151.865 | 0.0000 |
| 4,0" | 5,0 | 2151.595 | 2151.595 | 0.0001 |
| 5,0" | 6,0 | 2151.362 | 2151.362 | 0.0005 |
| 6,0" | 7,0 | 2151.043 | 2151.043 | 0.0000 |
| | | | | |

| Upper J, K e/f | Lower J, K e/f | Observed | Calculated | Obs - Calc |
|---|---|---|---|---|
| 4,0" | 3,1e | 2150.228 | 2150.229 | -0.0005 |
| 5,0" | 4,1e | 2150.295 | 2940.295 | -0.0003 |
| 6,0" | 5,1e | 2150.278 | 2150.276 | 0.0017 |
| 5,0" | 5,1f | 2149.427 | 2149.427 | -0.0003 |
| 6,0" | 6,1f | 2149.219 | 2149.219 | -0.0001 |

Table A-2. Observed transitions of CO-$O_2$ in group 2, correlating with ($n(O_2) = 1$, $j(O_2) = 2$) (units of cm$^{-1}$). Note: The "calculated" positions correspond to the "experimental" energy levels from Table 2 of the paper.

| Upper J, K e/f | Lower J, K e/f | Observed | Calculated | Obs - Calc |
|---|---|---|---|---|
| 3,3e | 2,2e | 2145.741 | 2145.741 | -0.0002 |
| 4,3e | 3,2e | 2145.891 | 2145.892 | -0.0011 |
| 5,3e | 4,2e | 2146.042 | 2146.043 | -0.0007 |
| 6,3e | 5,2e | 2146.196 | 2146.193 | 0.0027 |
| 3,3f | 2,2f | 2145.741 | 2145.741 | -0.0002 |
| 4,3f | 3,2f | 2145.894 | 2145.894 | 0.0007 |
| 5,3f | 4,2f | 2146.051 | 2146.050 | 0.0010 |
| 6,3f | 5,2f | 2146.212 | 2146.214 | -0.0023 |
| 3,3e | 3,2f | 2145.303 | 2145.303 | 0.0003 |
| 4,3e | 4,2f | 2145.309 | 2145.309 | -0.0002 |
| 5,3e | 5,2f | 2145.319 | 2145.319 | -0.0002 |
| 3,3f | 3,2e | 2145.299 | 2145.298 | 0.0007 |
| 4,3f | 4,2e | 2145.296 | 2145.295 | 0.0002 |
| 5,3f | 5,2e | 2145.288 | 2145.288 | -0.0005 |
| | | | | |
| 3,2e | 2,2e | 2143.136 | 2143.137 | -0.0004 |
| 4,2e | 3,2e | 2143.289 | 2143.289 | 0.0004 |
| 5,2e | 4,2e | 2143.447 | 2143.446 | 0.0012 |
| 6,2e | 5,2e | 2143.607 | 2143.608 | -0.0007 |
| 3,2f | 2,2f | 2143.132 | 2143.134 | -0.0014 |
| 4,2f | 3,2f | 2143.281 | 2143.281 | 0.0000 |
| 5,2f | 4,2f | 2143.431 | 2143.430 | 0.0011 |
| 6,2f | 5,2f | 2143.579 | 2143.580 | -0.0005 |
| 2,2e | 3,2e | 2142.253 | 2142.252 | 0.0006 |
| 3,2e | 4,2e | 2142.100 | 2142.100 | -0.0003 |
| 4,2e | 5,2e | 2141.943 | 2141.944 | -0.0005 |
| 5,2e | 6,2e | 2141.782 | 2141.782 | 0.0003 |
| 2,2f | 3,2f | 2142.256 | 2142.255 | 0.0009 |
| 3,2f | 4,2f | 2142.108 | 2142.107 | 0.0001 |
| 4,2f | 5,2f | 2141.958 | 2141.959 | -0.0012 |
| 5,2f | 6,2f | 2141.811 | 2141.810 | 0.0005 |
| | | | | |
| 2,2e | 3,3e | 2139.629 | 2139.629 | 0.0003 |
| 3,2e | 4,3e | 2139.481 | 2139.480 | 0.0008 |
| 4,2e | 5,3e | 2139.333 | 2139.332 | 0.0003 |
| 5,2e | 6,3e | 2139.185 | 2139.186 | -0.0018 |
| 2,2f | 3,3f | 2139.629 | 2139.628 | 0.0004 |
| 3,2f | 4,3f | 2139.478 | 2139.479 | -0.0008 |
| 4,2f | 5,3f | 2139.324 | 2139.325 | -0.0012 |
| 5,2f | 6,3f | 2139.168 | 2139.166 | 0.0025 |
| 3,2f | 3,3e | 2140.067 | 2140.067 | -0.0004 |
| 4,2f | 4,3e | 2140.064 | 2140.064 | 0.0000 |
| 5,2f | 5,3e | 2140.057 | 2140.057 | 0.0003 |
| 3,2e | 3,3f | 2140.071 | 2140.072 | -0.0004 |
| 4,2e | 4,3f | 2140.077 | 2140.077 | 0.0002 |
| | | | | |
| 1,1e | 2,2e | 2145.512 | 2145.510 | 0.0014 |
| 2,1e | 3,2e | 2145.360 | 2145.360 | -0.0002 |
| 3,1e | 4,2e | 2145.203 | 2145.203 | -0.0001 |
| 4,1e | 5,2e | 2145.041 | 2145.035 | 0.0062 |
| 1,1f | 2,2f | 2145.516 | 2145.516 | 0.0000 |
| 2,1f | 3,2f | 2145.376 | 2145.379 | -0.0022 |
| 3,1f | 4,2f | 2145.245 | 2145.244 | 0.0011 |
| 4,1f | 5,2f | 2145.115 | 2145.110 | 0.0051 |
| 3,1e | 2,2e | 2146.240 | 2146.240 | 0.0002 |
| 4,1e | 3,2e | 2146.387 | 2146.380 | 0.0069 |
| 3,1f | 2,2f | 2146.269 | 2146.270 | -0.0008 |
| 4,1f | 3,2f | 2146.439 | 2146.432 | 0.0067 |
| | | | | |
| 2,2e | 1,1e | 2139.859 | 2139.859 | -0.0003 |
| 3,2e | 2,1e | 2140.011 | 2140.011 | 0.0006 |
| 4,2e | 3,1e | 2140.169 | 2140.169 | -0.0002 |
| 2,2f | 1,1f | 2139.853 | 2139.854 | -0.0007 |
| 3,2f | 2,1f | 2139.995 | 2139.993 | 0.0022 |
| 4,2f | 3,1f | 2140.128 | 2140.130 | -0.0023 |
| 5,2f | 4,1f | 2140.268 | 2140.267 | 0.0008 |
| 2,2f | 2,1e | 2139.569 | 2139.567 | 0.0016 |
| 2,2e | 3,1e | 2139.132 | 2139.132 | -0.0003 |
| 3,2e | 4,1e | 2138.985 | 2138.994 | -0.0087 |
| 2,2f | 3,1f | 2139.103 | 2139.104 | -0.0006 |
| | | | | |
| 4,4e | 3,3e | 2149.849 | 2149.849 | 0.0004 |
| 5,4e | 4,3e | 2149.979 | 2149.978 | 0.0010 |
| 6,4e | 5,3e | 2150.104 | 2150.104 | 0.0007 |
| 7,4e | 6,3e | 2150.224 | 2150.224 | 0.0000 |
| 8,4e | 7,3e | 2150.341 | 2150.340 | 0.0014 |
| 4,4f | 3,3f | 2149.849 | 2149.850 | -0.0005 |
| 5,4f | 4,3f | 2149.979 | 2149.981 | -0.0012 |
| 6,4f | 5,3f | 2150.109 | 2150.108 | 0.0015 |
| 7,4f | 6,3f | 2150.228 | 2150.230 | -0.0021 |
| 8,4f | 7.3f | 2150.345 | 2150.346 | -0.0008 |
| 4,4e | 4,3f | 2149.261 | 2149.261 | 0.0002 |
| 5,4e | 5,3f | 2149.239 | 2149.240 | -0.0011 |
| 6,4e | 6,3f | 2149.219 | 2149.213 | 0.0061 |
| 4,4f | 4,3e | 2149.261 | 2149.258 | 0.0024 |
| 5,4f | 5,3e | 2149.236 | 2149.236 | 0.0006 |
| | | | | |
| 2,2'e | 2,2f | 2152.490 | 2152.491 | -0.0006 |
| 3,2'e | 3,2f | 2152.503 | 2152.502 | 0.0012 |
| 4,2'e | 4,2f | 2152.530 | 2152.529 | 0.0007 |
| 2,2'f | 2,2e | 2152.483 | 2152.483 | -0.0002 |
| 3,2'f | 3,2e | 2152.469 | 2152.470 | -0.0012 |
| 4,2'f | 4,2e | 2152.442 | 2152.443 | -0.0001 |
| 2,2'e | 3,2e | 2152.047 | 2152.048 | -0.0007 |
| 3,2'e | 4,2e | 2151.904 | 2151.903 | 0.0006 |
| 4,2'e | 5,2e | 2151.766 | 2151.767 | -0.0010 |
| 2,2'f | 3,2f | 2152.046 | 2152.045 | 0.0015 |
| 3,2'f | 4,2f | 2151.887 | 2151.888 | -0.0006 |
| 4,2'f | 5,2f | 2151.719 | 2151.719 | 0.0001 |
| 4,2'e | 3,2e | 2153.112 | 2153.112 | -0.0002 |
| 4,2'f | 3,2f | 2153.041 | 2153.041 | 0.0006 |

Table A-3. Calculated energy levels of CO-O$_2$ (in cm$^{-1}$) with $J = 0$ to 5, relative to the zero point energy (ZPE) of -81.9332 cm$^{-1}$. This ZPE corresponds to the forbidden level with $J = 0$, $K = 0$, $n(O_2) = 0$. Stack labels A, B, C, etc. correspond to those of Tables 5 and 6 in the paper. Permutation inversion group symmetry and spectroscopy parity $e/f$ are also indicated. Only the allowed B+ and B− levels are given.

| $J = 0, B^+(e)$ | | | $J = 0, B^-(f)$ | | |
|---|---|---|---|---|---|
| calc | K | stack | calc | K | stack |
| 4.6342 | 0 | C | 4.4928 | 0 | B |
| 8.3172 | 0 | E | 16.7199 | 0 | |
| 16.2495 | 0 | | | | |

| $J = 1, B^+(f)$ | | | $J = 1, B^-(e)$ | | |
|---|---|---|---|---|---|
| calc | K | stack | calc | K | stack |
| 1.8396 | 1 | A | 1.8322 | 1 | A |
| 4.6507 | 0 | B | 4.7915 | 0 | C |
| 9.6674 | 1 | F | 8.4632 | 0 | E |
| 11.6413 | 1 | G | 9.6652 | 1 | F |
| 11.8904 | 1 | I | 11.6342 | 1 | G |
| 16.3491 | 1 | | 11.8904 | 1 | I |
| 16.9134 | 0 | | 16.2400 | 1 + 0 | |
| | | | 16.5291 | 0 + 1 | |

| $J = 2, B^+(e)$ | | | $J = 2, B^-(f)$ | | |
|---|---|---|---|---|---|
| calc | K | stack | calc | K | stack |
| 2.1410 | 1 | A | 2.1634 | 1 | A |
| 4.8824 | 2 | D | 4.8841 | 2 | D |
| 5.1076 | 0 | C | 4.9665 | 0 | B |
| 8.7552 | 0 | E | 9.9743 | 1 | F |
| 9.9676 | 1 | F | 11.9416 | 1 | G |
| 11.9207 | 1 | G | 12.1993 | 1 | I |
| 12.1989 | 1 | I | 15.4010 | 2 | K |
| 15.3988 | 2 | K | 16.5891 | 1 | |
| 16.4259 | 1 + 0 | | 17.2837 | 0 | |
| 16.9096 | 0 + 1 | | 17.6192 | 2 | |
| 17.6191 | 2 | | 18.5848 | 2 | |
| 18.5843 | 2 | | | | |

| J = 3, B⁺(f) | | | J = 3, B⁻(e) | | |
|---|---|---|---|---|---|
| calc | K | stack | calc | K | stack |
| 2.6489 | 1 | A | 2.6040 | 1 | A |
| 5.3533 | 2 | D | 5.3455 | 2 | D |
| 5.4400 | 0 | B | 5.5853 | 0 | C |
| 10.4344 | 1 | F | 9.1930 | 0 | E |
| 12.0063 | 3 | H | 10.4206 | 1 | F |
| 12.3920 | 1 | G | 12.0062 | 3 | H |
| 12.6631 | 1 | I | 12.3514 | 1 | G |
| 14.8465 | 3 | J | 12.6617 | 1 | I |
| 15.8341 | 2 | K | 14.8465 | 3 | J |
| 16.9714 | 1 | | 15.8232 | 2 | K |
| 17.8147 | 0 | | 16.7708 | 1 + 0 | |
| 18.0735 | 2 | | 17.4179 | 0 + 1 | |

| J = 4, B⁺(e) | | | J = 4, B⁻(f) | | |
|---|---|---|---|---|---|
| calc | K | stack | calc | K | stack |
| 3.2210 | 1 | A | 3.2958 | 1 | A |
| 5.9565 | 2 | D | 5.9784 | 2 | D |
| 6.2278 | 0 | C | 6.0709 | 0 | B |
| 9.7765 | 0 | E | 11.0475 | 1 | F |
| 11.0238 | 1 | F | 12.6075 | 3 | H |
| 12.6066 | 3 | H | 12.9924 | 1 | G |
| 12.9277 | 1 | G | 13.2825 | 1 | I |
| 13.2787 | 1 | I | 15.4392 | 3 | J |
| 15.4386 | 3 | J | 16.4092 | 2 | K |
| 16.3772 | 2 | K | 17.5035 | 1 | |
| 17.2811 | 1 + 0 | | 18.4947 | 0 | |
| 18.0481 | 0 + 1 | | 18.0718 | 4 | |

|  | $J = 5$, $B^+(f)$ |  |  | $J = 5$, $B^-(e)$ |  |
|---|---|---|---|---|---|
| calc | K | stack | calc | K | stack |
| 4.1037  | 1 | A | 3.9914  | 1 | A |
| 6.7588  | 2 | D | 6.7128  | 2 | D |
| 6.8588  | 0 | B | 7.0367  | 0 | C |
| 11.8129 | 1 | F | 10.5055 | 0 | E |
| 13.3553 | 3 | H | 11.7761 | 1 | F |
| 13.7425 | 1 | G | 13.3518 | 3 | H |
| 14.0577 | 1 | I | 13.6515 | 1 | G |
| 16.1796 | 3 | J | 14.0498 | 1 | I |
| 17.1249 | 2 | K | 16.1770 | 3 | J |
| 18.1882 | 1 |   | 17.0549 | 2 | K |
| 18.8435 | 4 |   | 17.9622 | 1 + 0 |   |
| 19.3113 | 0 |   | 18.7969 | 0 + 1 |   |
| 19.4490 | 2 |   | 18.8535 | 4 |   |